\documentclass[aps,pra,reprint,superscriptaddress,floatfix]{revtex4-2}

\usepackage{amsfonts}
\usepackage{amsmath}
\usepackage{amssymb}
\usepackage{graphicx}
\usepackage{color}
\usepackage{physics}
\usepackage{float}
\usepackage{mathtools}
\usepackage{algorithm}
\usepackage{algpseudocode}
\usepackage{wrapfig}
\usepackage{mathbbol}
\usepackage{upgreek}
\usepackage{subcaption}
\usepackage[normalem]{ulem}

\usepackage[colorinlistoftodos]{todonotes}
\usepackage[colorlinks=true, allcolors=blue]{hyperref}
\usepackage{hyperref}

\usepackage{mwe}
\usepackage{epstopdf}
\usepackage[doipre={DOI:~}]{uri}

\newcommand \be {\begin{eqnarray}}
\newcommand \ee {\end{eqnarray}}
\newcommand \ben {\begin{eqnarray}}
\newcommand \een {\end{eqnarray}}

\begin{document}
\title{Topological interactions in vortex-wave collisions in Bose-Einstein condensates}
\author{Vebjørn Øvereng}
\affiliation{Departament de F\'{i}sica, Universitat Polit\`{e}cnica de Catalunya, Campus Nord B4-B5, E-08034 Barcelona, Spain}
\affiliation{Njord Centre, Department of Physics, University of Oslo, 0371 Oslo, Norway}
\author{Andrew Baggaley}
\affiliation{Joint Quantum Centre (JQC) Durham-Newcastle, School of Mathematics Statistics and Physics,
Newcastle University, Newcastle upon Tyne, NE1 7RU, UK}
\affiliation{Department of Mathematics and Statistics, Lancaster University, Lancaster, LA1 4YF, UK}
\author{Luiza Angheluta}
\affiliation{Njord Centre, Department of Physics, University of Oslo, 0371 Oslo, Norway}
\email{luiza.angheluta@fys.uio.no}

\begin{abstract} 
We study vortex-vortex and vortex-wave collisions in two-dimensional weakly interacting Bose-Einstein condensates, processes that play a central role in decaying quantum turbulence.  Using numerical simulations of the Gross-Pitaevskii equation, we show that during collisions of vortex-antivortex dipoles, the kinetic energy is transferred from incompressible to compressible modes by two distinct mechanisms. Below the critical vortex separation for annihilation, the transfer is mediated by quantum energy released during annihilation events, while above the threshold it arises from vortex acceleration. In wave-vortex collisions, an incoming solitary wave splits into transient phase slips that interact with the vortex, one of the phase slips contributes to vortex annihilation, and the other phase slip acquires a stable core and forms a new vortex. By analyzing vortex trajectories and energy spectra, we provide new insights into energy transfer mechanisms in quantum turbulence and offer broader implications for topological interactions mediated by vortices.
\end{abstract}
\maketitle

In quantum fluids, vorticity is localized at the nodal points
 of the complex wavefunction $\psi$ in two dimensions (or nodal lines in three dimensions). Around these points,  where the condensate density vanishes $\|\psi\| = 0$, the condensate phase  $\arg(\psi)$ winds by integer multiples of $2\pi$, giving rise to quantum vortices that carry quantized circulation. These quantum vortices are  fundamental building blocks of superfluid flows and play a central role in the emergence of quantum turbulence~\cite{barenghi2014introductiontoQT, white_vortices_2014,  bradley_energy_2012}. 

In two-dimensional Bose–Einstein condensates (BECs), vortex–vortex collisions are considered as primary dissipative mechanism contributing to the decay of quantum turbulence~\cite{kwon_relaxation_2014, stagg_generation_2015, groszek_onsager_2016, valani_einsteinbose_2018, baggaley_decay_2018}. For  developing a comprehensive picture of quantum turbulence decay, there is still an open challenge to understand the processes involved in these vortex–vortex collisions and vortex–wave interactions. While the dynamics and interactions between vortices have been extensively studied, e.g., \cite{chaves2011vortex, fetter2001vortices}, the influence of solitary waves in topological collision processes remains comparatively underexplored, despite its critical importance for mechanisms of energy transfer, vortex annihilation, and creation. 

We investigate vortex-vortex collisions and vortex-wave interactions in weakly interacting, two-dimensional BECs by performing numerical simulations of the Gross–Pitaevskii equation. Specifically, we focus on two scenarios: i) collisions between solitary waves and single vortices, which reveal rich topological dynamics involving the transient formation of phase slips \autoref{fig:exchange}, leading to annihilation of the vortex with the phase slip of opposite circulation, and the nucleation of a new vortex; ii) head-on collisions between vortex dipoles \autoref{fig:collision}, where we uncover two distinct energy-transfer mechanisms that operate above and below the critical vortex separation for annihilation. By analyzing vortex trajectories and energy spectra, we characterize the mechanisms by which incompressible energy is converted into compressible energy, and find that the quantum pressure mediates this transfer. These results highlight the role of defect dynamics in driving energy cascades and decay of 2D quantum turbulence.

By connecting these dissipative topological processes to the broader context of decaying quantum turbulence, we bridge the gap between vortex interactions and macroscopic turbulent behavior in two-dimensional quantum fluids. Analogous dissipative mechanism mediated by vortex-wave interactions are likely at play in diverse physical systems, including superconductors and neutron star interiors. 

The paper is organized as follows. In Section~\ref{Sec:GPE}, we present the Gross–Pitaevskii theory and a non-singular field theory for tracking vortices and solitary waves~\cite{skogvoll_unified_2023} based on the vorticity of the superfluid current, and the energy decomposition used in our analysis. In Section~\ref{Sec:vortex-wave}, we discuss the scattering of a solitary wave from a single vortex and the associated topological exchange processes. Section~\ref{Sec:4-vortex} focuses on four-vortex dynamics during dipole–dipole collisions, highlighting the thresholds for partner exchange versus annihilation and comparing results with point-vortex predictions. Section~\ref{Sec:decay_turb} connects these elementary processes to the regime of decaying quantum turbulence, analyzing vortex number decay and the transfer of energy between incompressible, compressible, and quantum components. Final conclusions are presented in Section~\ref{Sec:conclusion}.

\section{Vortices in the Gross-Pitaevski theory}\label{Sec:GPE}

We model the Bose-Einstein condensate at finite temperature using the damped Gross-Pitaevskii equation~\cite{madeira_quantum_2020_AVS, baggaley_decay_2018, bradley_energy_2012, ronning_nucleation_2022, skaugen_universal_2016}
\begin{multline}
    i \hbar\partial_t \Psi(t, \mathbf{r}) =\\
    (1 - i\gamma)\left(-\frac{\hbar^2}{2m}\nabla^2 + V_{\text{Ext}}(\mathbf{r}) - \mu + g|\Psi(\mathbf{r})|^2\right) \Psi(\mathbf{r}),\label{eq:GPE}
\end{multline}
where $\Psi(t, \mathbf{r})$ is the order parameter of the condensate, $V_{\text{Ext}}(\mathbf{r})$ is the trapping potential, $\mu$ the chemical potential of the condensate, $m$ the particle mass, $g$ sets the interaction strength, and $\gamma$ is a phenomenological damping coefficient representing the dissipative effects that arise from coupling to a thermal reservoir.  

The wavefunction
$\Psi = \sqrt{\rho}\exp(i\theta)$ in the Madelung transformation, encodes both the condensate density, $\rho = \|\Psi\|^2$ and also the phase, $\theta = \arg(\Psi)$. The latter determines the superfluid flow through the current density, $\mathbf J = \frac{\hbar}{m}\rho \nabla \theta= \frac{\hbar}{m}\Im(\Psi^*\nabla\Psi) $, which directly links phase gradients to observable transport properties~\cite{madeira_quantum_2020_AVS, ronning_nucleation_2022}. Quantum vortices appear as topological defects where the wavefunction vanishes in magnitude, ensuring single-valuedness of $\Psi$ across the $2\pi$-phase windings.
We locate vortices using the pseudo-vorticity (or the $D$-field), defined as the curl of the superfluid current
\cite{villois2016vortex, ronning_nucleation_2022, skogvoll_unified_2023}
\begin{equation}
    D = \frac{1}{2} (\nabla\times \mathbf{J})_z = \frac{1}{2}\varepsilon_{ij}\partial_i \Psi^* \partial_j \Psi. 
\end{equation}
The $D$-field has several useful properties. Unlike the condensate phase $\theta$, which is undetermined at topological defects and discontinuous at the associated branch cuts, 
the $D$ field  is a smooth scalar field that can be computed directly from the wavefunction without unwrapping the phase. It vanishes in regions of nearly uniform condensate phase and is localized at the cores of quantized vortices and other high-energy excitations such as solitons~\cite{ronning_nucleation_2022, skogvoll_unified_2023}. This strong spatial localization makes the $D$-field a powerful tool for identifying and tracking defects, as vortex positions coincide with well-defined extrema. The sign of the $D$-field encodes the sign of circulation, allowing for straightforward distinction between vortices and antivortices~\cite{ronning_nucleation_2022}.

\subsection{Hydrodynamic energy}
For analyzing the kinetic energy, it is convenient to introduce the density-weighted velocity $\mathbf{u} = \sqrt{\rho}\mathbf{v} = \frac{\hbar}{m}\sqrt{\rho}\nabla\theta$,
such that the total kinetic energy of the condensate can be expressed as \cite{bradley_energy_2012, bradley_spectral_2022}
\begin{align}
    E_{kin} &= \frac{\hbar^2}{2m}\int \dd \mathbf{r} \: |\nabla\Psi|^2\nonumber\\
    &= \frac{m}{2} \int \dd \mathbf{r} \: |\mathbf{u}|^2 + \frac{\hbar}{2m}\int \dd \mathbf{r} \: |\nabla\sqrt{\rho}|^
2.
\end{align}
The second term, often referred to as the quantum energy, $E_q$, originates from steep density gradients in the condensate. It is most pronounced within the vortex core profile and near the edges of the condensate where the density drops at the trap boundary.

Applying a Helmholtz decomposition, the $\mathbf{u}$-field can be separated into a compressible part, $\mathbf{u}^c$ that is curl-free, and an incompressible part, $\mathbf{u}^i$ that is divergence-free. This leads to a decomposition of the kinetic energy into three contributions \cite{bradley_energy_2012}
\begin{align}
    E_i =& \frac{m}{2}\int \dd \mathbf{r} \: |\mathbf{u}^i|^2 \\
    E_c =& \frac{m}{2}\int \dd \mathbf{r} \: |\mathbf{u}^c|^2 \\
    E_q =& \frac{\hbar^2}{2m}\int \dd \mathbf{r} \: |\nabla\sqrt{\rho}|^2.
\end{align}
The incompressible part, $E_i$, is associated with the rotational superfluid velocity around quantized vortices and is determined primarily by their spatial configuration. The compressible part, $E_c$, corresponds to density fluctuations and is related to the propagation of sound waves in the condensate. The balance between these two contributions, $E_i$ and $E_c$, provides a useful framework for characterizing the interaction between vortices and collective wave excitations. 

\section{Vortex-wave interaction}\label{Sec:vortex-wave}

At zero temperature, a vortex–antivortex pair forms a stable dipole that drifts through the condensate~\cite{groszek_onsager_2016}. Direct annihilation in this regime requires many-body collisions, most notably a four-vortex process ~\cite{kanai_true_2021, groszek_onsager_2016}. In the presence of sound waves, a three-body process is also possible~\cite{kanai2024dynamical}. The four-vortex annihilation is thought to proceed in two stages: first, a vortex–antivortex pair collapses under the influence of a third vortex, forming a solitary wave; second, this wave collides with a fourth vortex, converting its energy into sound. As we show below, the outcome of the second stage depends sensitively on the energy of the solitary wave.

To probe this second stage, we imprint two vortices in close proximity and let them annihilate under real-time evolution using Eq.~\ref{eq:GPE}, producing a solitary wave that collides with a single vortex. As the solitary wave approaches the vortex, a pair of quasi-defects with opposite circulation emerges within it. These quasi-defects appear as phase singularities (phase slips) but lack the defect core, and thus do not form stable vortices~\cite{ronning_nucleation_2022}. The negative quasi-defect annihilates with the vortex, while the positive quasi-defect subsequently develops a stable core. This process constitutes a topological exchange, which can be separated into three stages defined by the number of phase slips in the interaction region: i) \emph{Approach:} only the original vortex is present; as the solitary wave approaches interaction with the vortex increases its pseudo-vorticity and incompressible kinetic energy. ii) \emph{Exchange:} three phase slips coexist as the negative quasi-defect moves toward the vortex. iii) \emph{Aftermath:} the negative quasi-defect annihilates with the vortex, producing a new outgoing wave.

\begin{figure}
\centering
    \includegraphics[width=\linewidth]{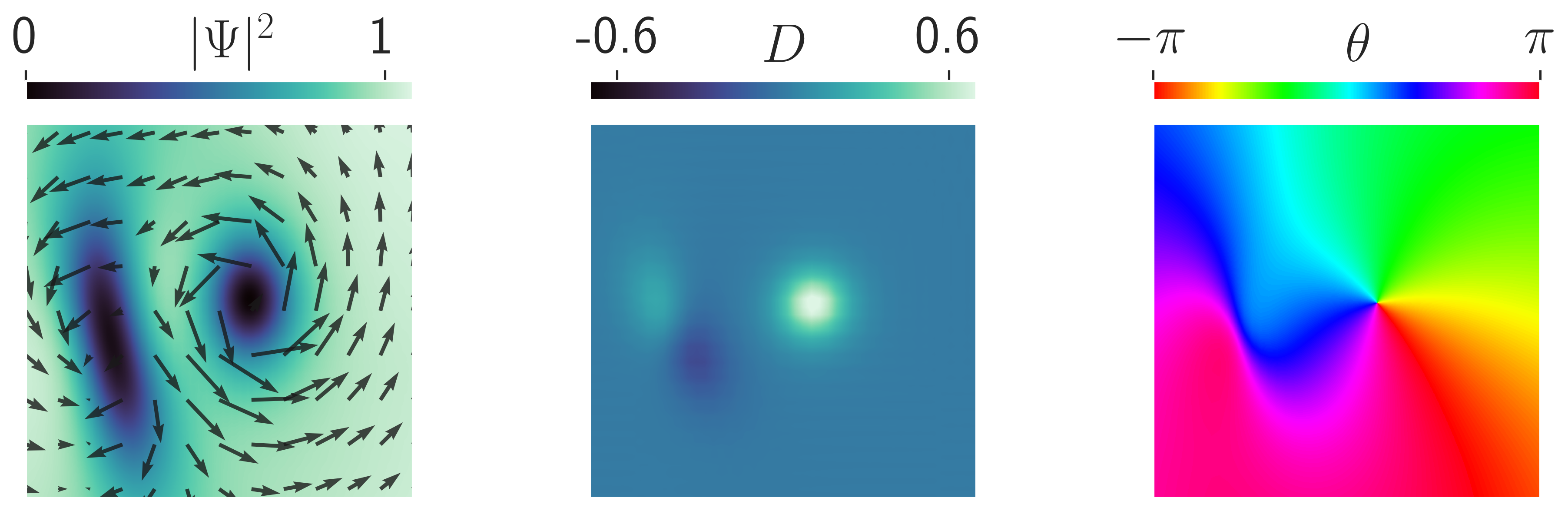}
    \caption{\textbf{Approach of the solitary wave to a vortex.} Left panel: condensate density $|\Psi|^2$ with the superfluid current $\mathbf J$. Middle panel: pseudo-vorticity ($D$-field). Right panel: condensate phase $\theta$. The solitary wave, generated by imprinting with a vortex-antivortex pair with initial separation of $\Delta r = 1.6\xi$, carries weak pseudo-vorticity and is rotated so that its lobe with negative pseudo-vorticity approaches the vortex.} 
        \label{fig:approach}
\end{figure}

\begin{figure}
    \centering
    \includegraphics[width=\linewidth]{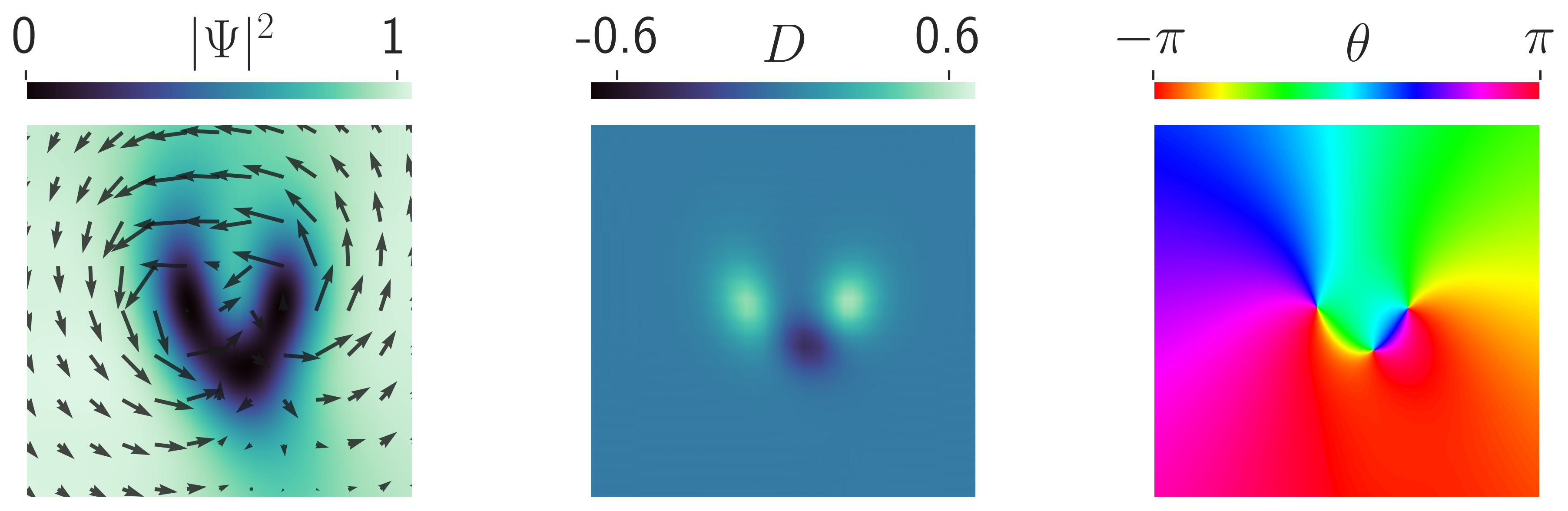}
    \caption{\textbf{Topological exchange during the collision.} Left panel: condensate density $|\Psi|^2$ with superfluid current $\mathbf J$ as vector field. Middle panel: pseudo-vorticity, $D$-field. Right panel: condensate phase $\theta$.}
    \label{fig:exchange}
\end{figure}

\begin{figure}
    \centering
    \includegraphics[width=\linewidth]{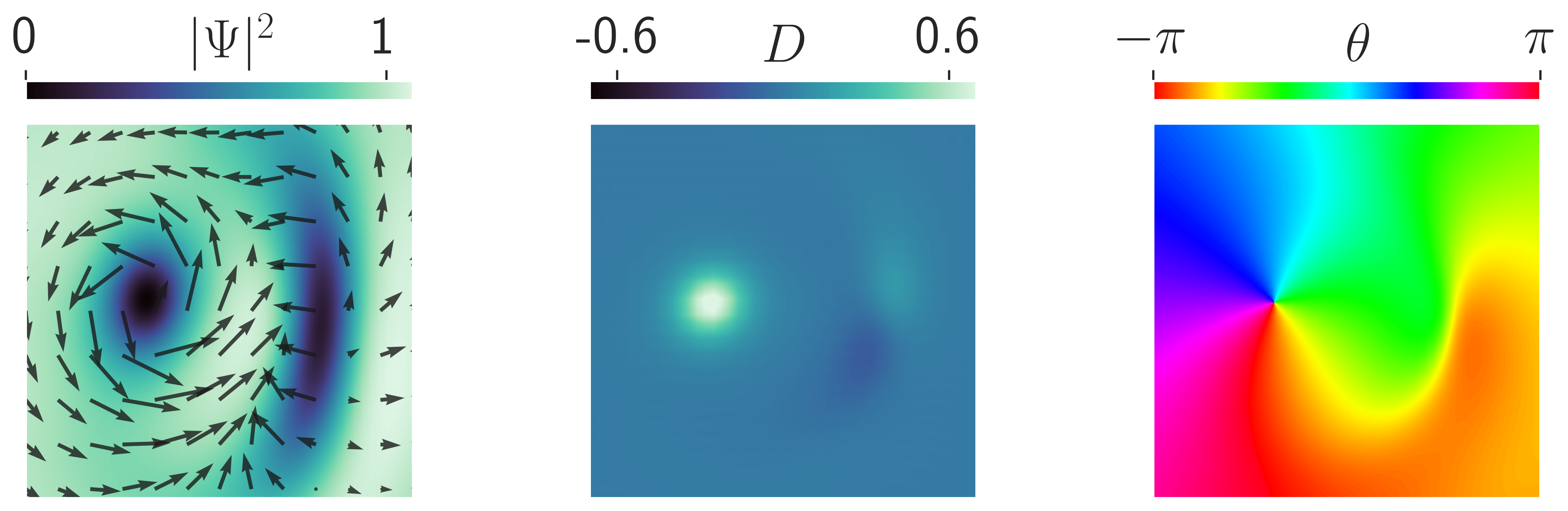}
    \caption{\textbf{After the exchange.} Left panel: condensate density $|\Psi|^2$ with superfluid current $\mathbf J$. Middle panel: pseudo-vorticity, $D$-field. Right panel: condensate phase $\theta$.}   
    \label{fig:post_collision}
\end{figure}

\autoref{fig:approach} shows a snapshot during the approach. The solitary wave carries nonzero pseudo-vorticity with localised regions with positive and negative circulation, which are rotated such that the negative pseudo-vorticity lobe moves toward the vortex.  At this stage, the only phase slips is that of the vortex. During the exchange (\autoref{fig:exchange}), the condensate density between the solitary wave and vortex drops, forming a continuous trench. The pseudo-vorticity strengthens markedly compared with the approach stage, and two additional phase slips appear in the condensate phase. After the exchange (\autoref{fig:post_collision}), the outgoing wave is rotated by the vortex and carries reduced pseudo-vorticity and incompressible kinetic energy compared with the incoming wave.

\begin{figure}
    \centering
    \includegraphics[width=\linewidth]{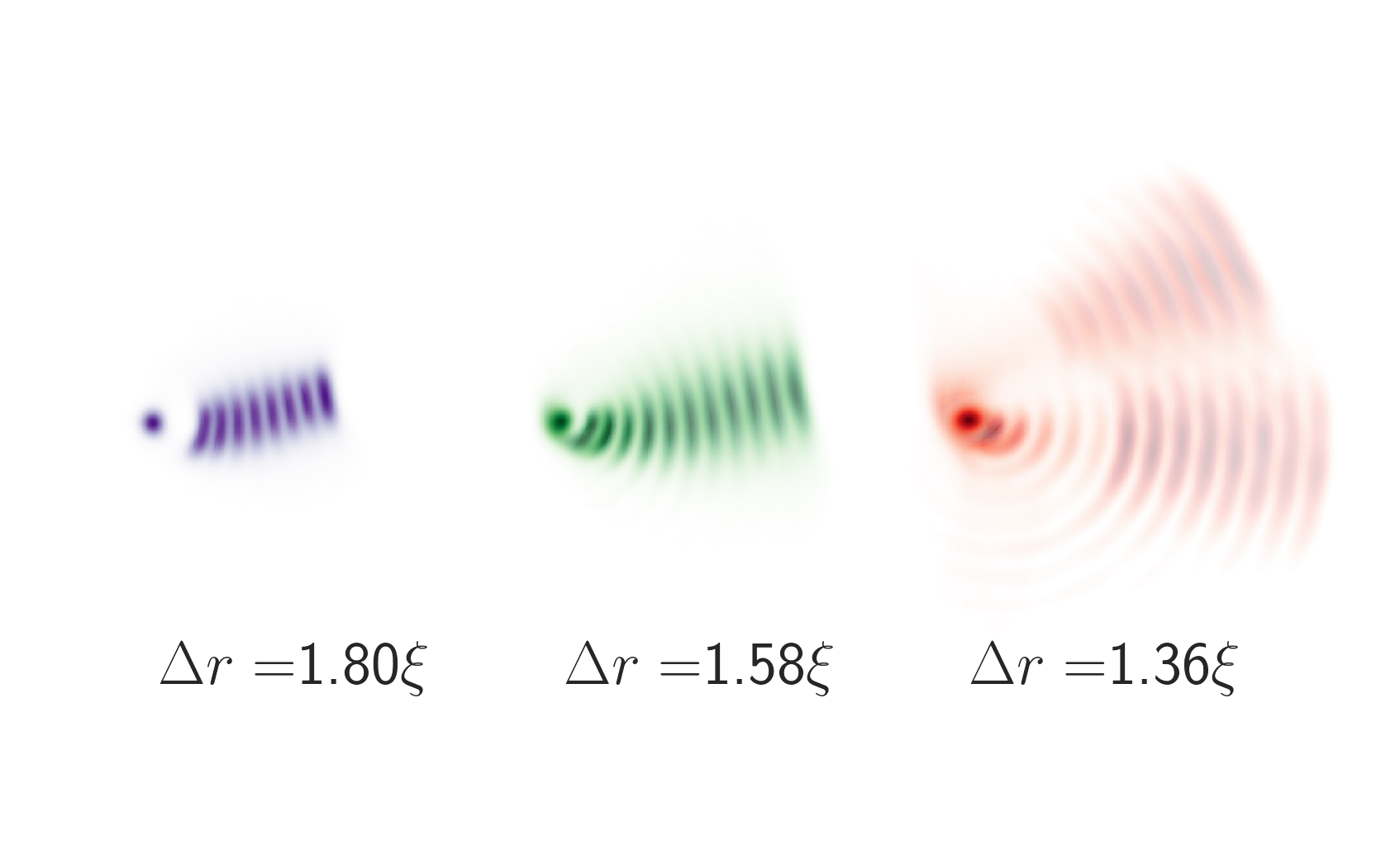}
    \caption{\textbf{Evolution of the outgoing solitary wave after the exchange.} Every $2\tau$ we plot the region with the largest area where $\rho < 0.9\rho_0$. The figure shows the motion of a single solitary wave over time. Colours indicate different initial dipole separations. The vividness of colour indicates the depth of the wave. Larger separations (left) produce compact solitary waves that persist, while smaller separations (right) yield low-energy waves that rapidly disperse into sound.}
    \label{fig:wave_after_collision}\end{figure}

\autoref{fig:wave_after_collision} shows the subsequent propagation of outgoing solitary waves generated with different initial dipole separations. For small separations (right), the initial solitary wave has low pseudo-vorticity and disperses rapidly into sound after the collision. For larger separations (left), the solitary wave is more dipole-like and retains its integrity after the collision, sometimes persisting as a shape-preserving solitary wave. In certain conditions, such waves can even reconstitute into vortices through further interactions with other vortices or with the trap boundary, a behavior we have also observed in turbulent simulations.

Thus, the second stage of the four-vortex annihilation process depends crucially on the energy of the solitary wave. Low-energy solitary waves readily dissipate into sound, while higher-energy waves can survive the collision and continue to participate in vortex dynamics.

\section{Vortex-vortex collisions}\label{Sec:4-vortex}
\begin{figure}[htbp]
\begin{minipage}{0.85\linewidth}
\includegraphics[width=\linewidth]{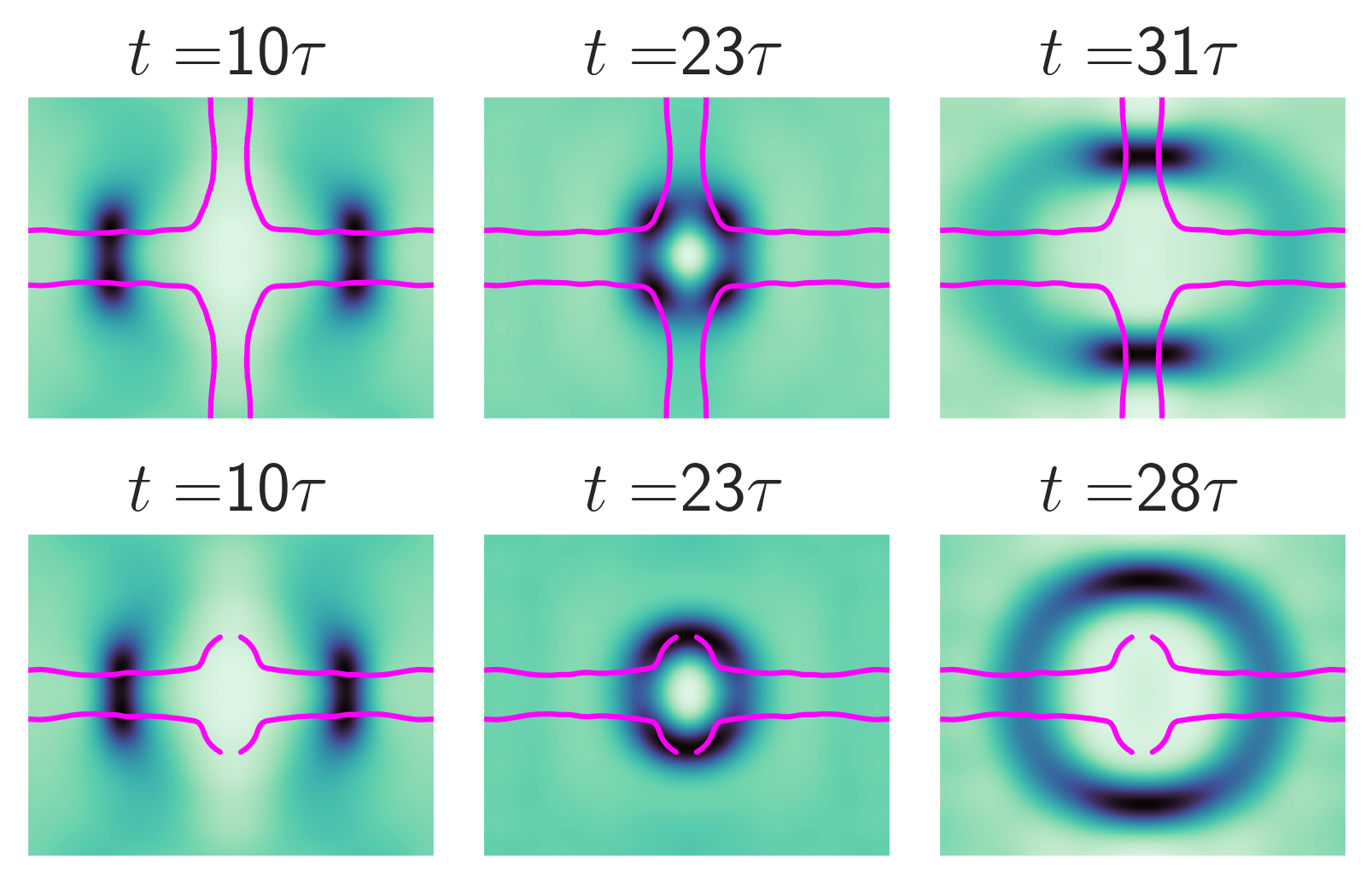}
\end{minipage}%
\begin{minipage}{0.1\linewidth}
\vspace{0.12cm}
\includegraphics[height = 4cm]{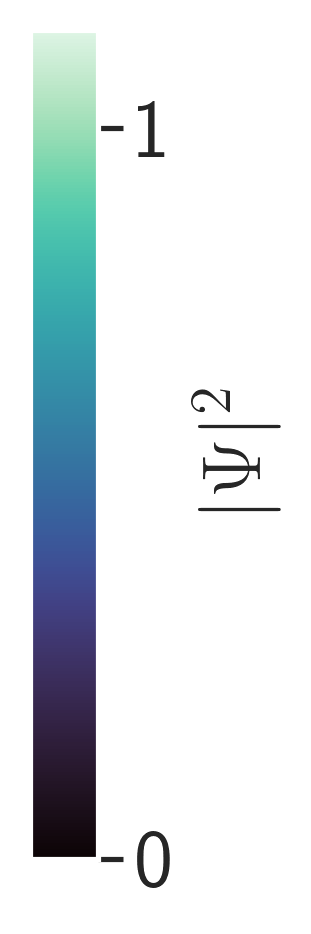}
\end{minipage}
\caption{\textbf{Evolution of the condensate density during dipole collision.} Top panel: dipole collision with initial separation $\Delta r = 2.5\xi$. Bottom panel: $\Delta r = 2.1\xi$. Pink lines trace the phase slip trajectories. For $\Delta r < \Delta r_c$ (bottom) the slips annihilate, while for $\Delta r > \Delta r_c$ (top) they survive the collision, and the outgoing dipoles stabilize at a new separation.}
\label{fig:collision}
\end{figure}

An alternative way to probe four-vortex annihilation is to set up two dipoles approaching each other head-on. The resulting dynamics are illustrated in \autoref{fig:collision}. When the initial dipole separation is smaller than a critical value, $\Delta r_c \approx 2.4\xi$, the vortices annihilate: the phase slips overlap and vanish, releasing a solitary wave (lower panels). For larger separations, $\Delta r > \Delta r_c$, the dipole is stable, and annihilation does not occur. Instead, the vortices exchange partners, forming new dipoles that propagate apart (upper panels).

\begin{figure}
\centering
\includegraphics[width=\linewidth]{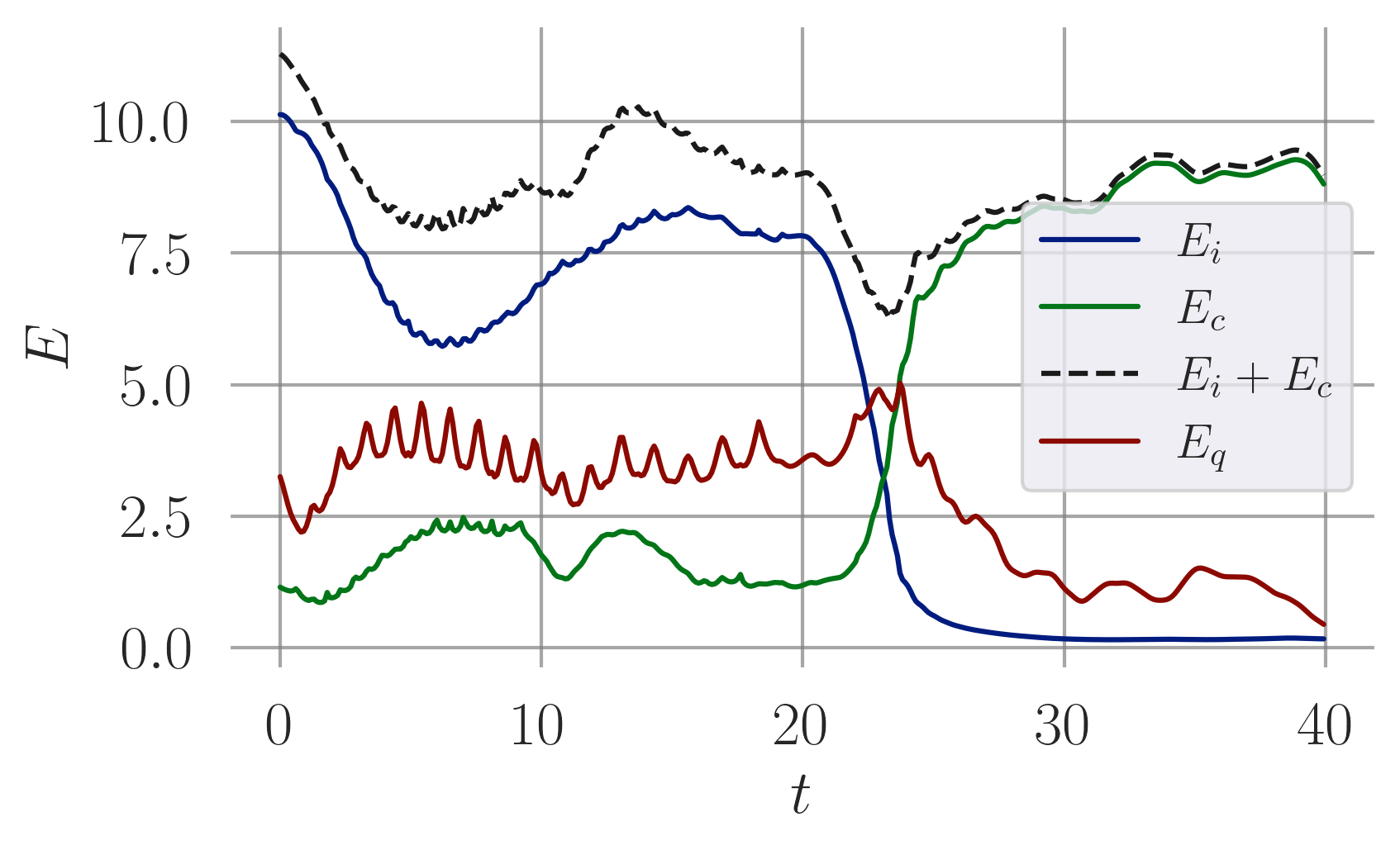}
\caption{Kinetic and quantum energy evolution during annihilation of dipoles with initial separation $\Delta r = 2.1\xi$. Between $t=20\tau$ and $t=25\tau$, $E_i$ drops sharply while $E_c$ rises. The quantum energy $E_q$ peaks during the event, before relaxing to a lower level than initially. $E_q$ is shifted down by $25\mu_0$ for clarity.}
\label{fig:dipdip_hydro_energy_annihilation}
\end{figure}

The energy evolution below $\Delta r_c$ is shown in \autoref{fig:dipdip_hydro_energy_annihilation}. Initially, the incompressible kinetic energy $E_i$ dominates, while the compressible part $E_c$ is small. The vortex cores and trap edges contribute to the quantum energy $E_q$, which peaks sharply during annihilation. This peak coincides with the drop in $E_i + E_c$, suggesting that the transfer of energy is not direct, but mediated through the quantum energy: $E_i \rightarrow E_q \rightarrow E_c$.

\begin{figure}
\centering
\includegraphics[scale = 0.5]{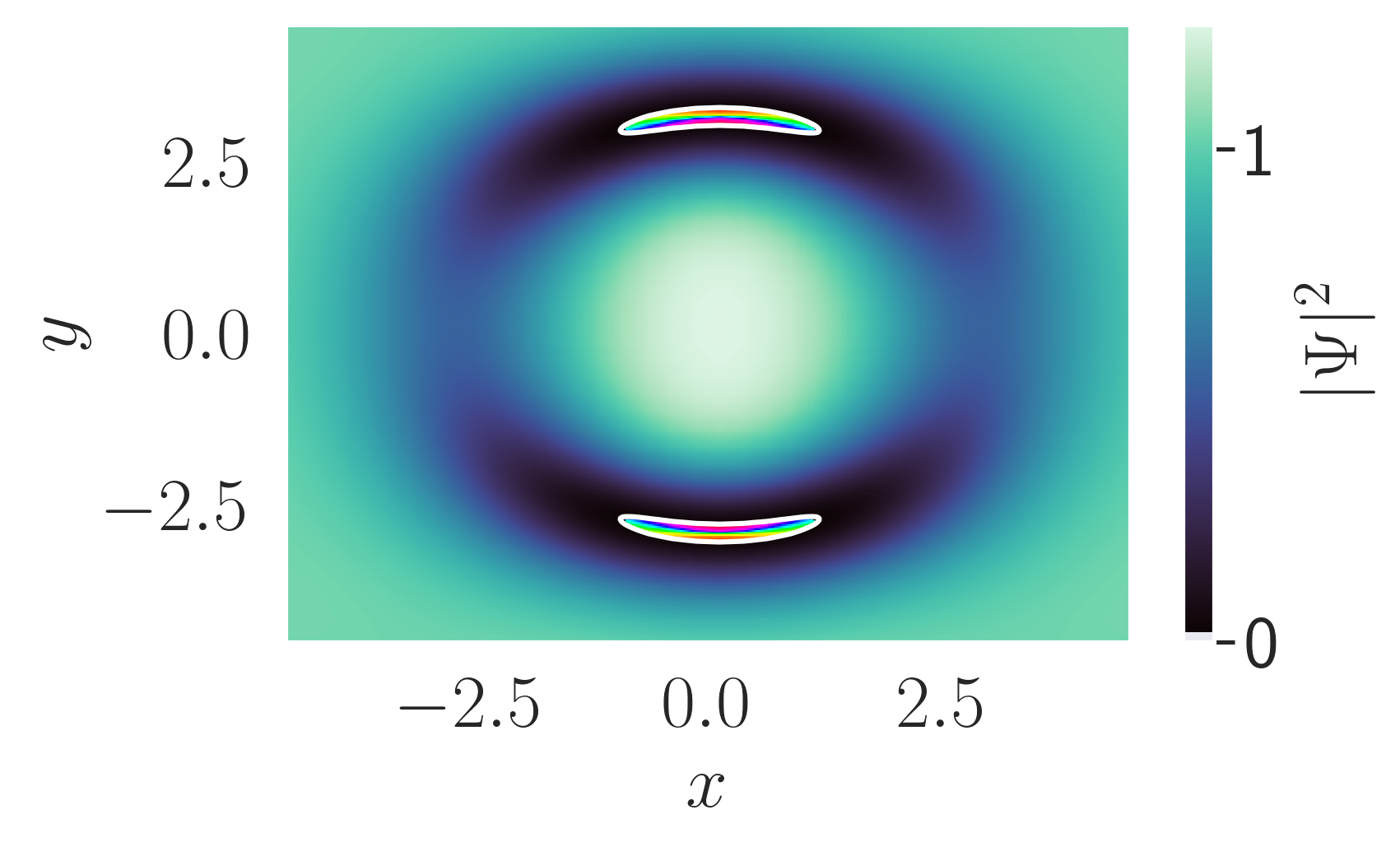}
\caption{Condensate density at $t=23.3\tau$ for $\Delta r = 2.1\xi$, corresponding to the peak in $E_q$ (cf. \autoref{fig:dipdip_hydro_energy_annihilation}). The phase is plotted within the low-density region $\rho < 0.03\rho_0$. Vortex cores are typically found for $\rho < 0.02\rho_0$; here the extended core region spans an area of $A\approx\xi^2$, compared to $A\approx0.1\xi^2$ for a single vortex.}
\label{fig:dipdip_max_Eq}
\end{figure}

At the peak of $E_q$, the condensate density (\autoref{fig:dipdip_max_Eq}) has an extended low-density region, far larger than the core of a single vortex. The surrounding gradients account for the increase in $E_q$ during annihilation.

\begin{figure}
\centering
\includegraphics[width = \linewidth]{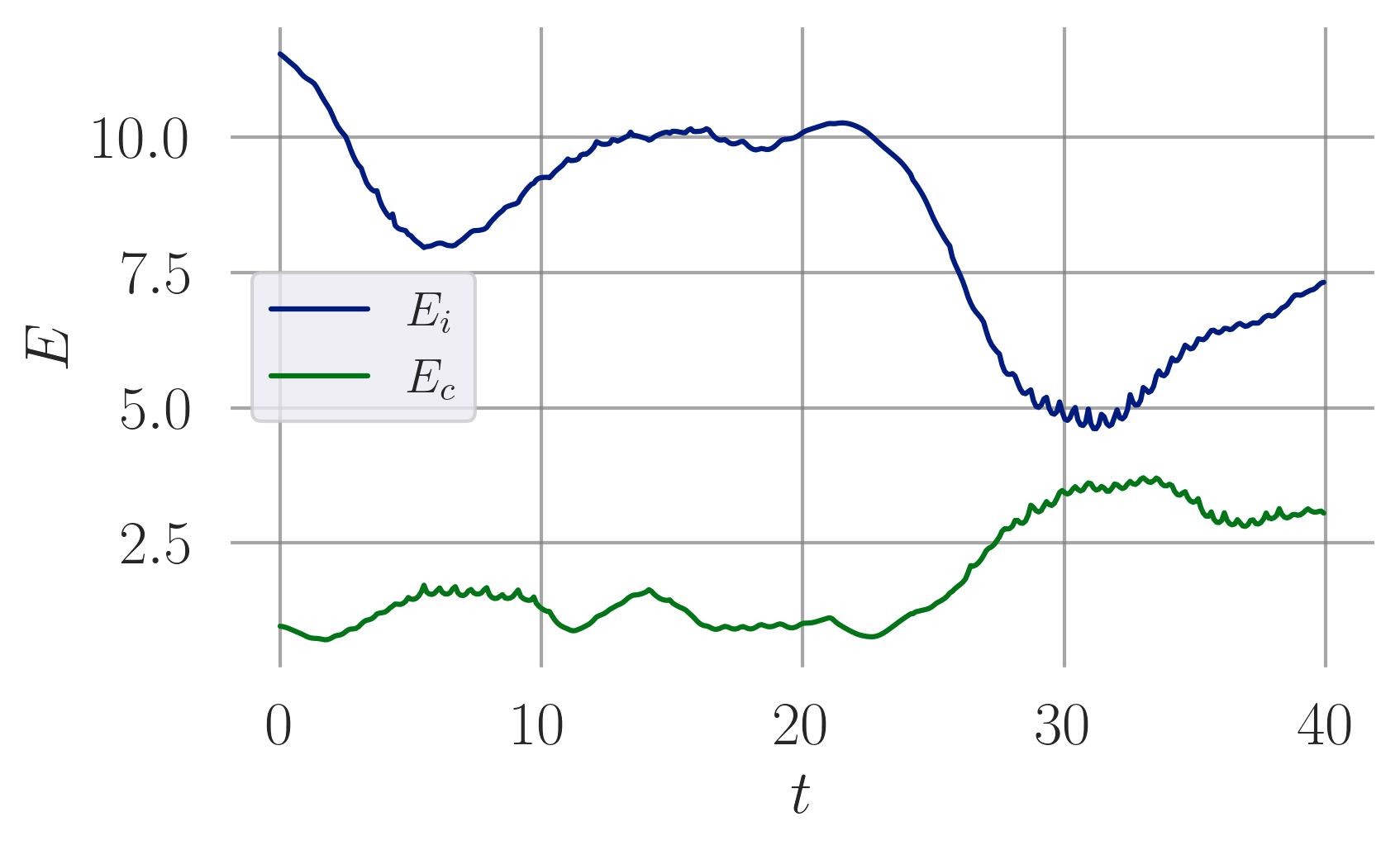}
\caption{Kinetic energy evolution for dipoles with initial separation $\Delta r = 2.5\xi$. During the collision $E_i$ decreases while $E_c$ rises. Afterwards $E_i$ gradually increases as the outgoing vortices separate.}
\label{fig:dipdip_hydro_energy_collision}
\end{figure}

For initial separations just above $\Delta r_c$, such as $\Delta r = 2.5\xi$, the incompressible energy $E_i$ decreases during the collision but quickly recovers (\autoref{fig:dipdip_hydro_energy_collision}). This increase in $E_i$ corresponds to the vortices separating further in the outgoing dipoles, accompanied by a reduction in $E_c$. The effect arises from overlap of the vortex density profiles and disappears when the vortices are well separated. In that regime the dynamics can be captured by the point vortex model in a homogeneous condensate \cite{parker_controlled_2004, ronning_nucleation_2022, skaugen_universal_2016}.

For large initial separations, $\Delta r \gg \Delta r_c$, the vortices interact only weakly through their density profiles, and the dynamics reduce to those of point vortices in a homogeneous condensate \cite{parker_controlled_2004, ronning_nucleation_2022, skaugen_universal_2016}. In this limit the analogy to accelerating charges becomes useful: just as charges radiate electromagnetic waves, accelerating vortices radiate sound \cite{simula_gravitational_2020}. By the Larmor formula, the radiated energy is proportional to the square of the acceleration, giving a scaling $E_c\sim\int dt a^2$~\cite{parker_controlled_2004, kwon_sound_2021}. 

For a head-on collision where the vortices deflect by approximately $90^\circ$, the collision time scales as $\Delta t \sim \Delta r / v$, with vortex velocity $v \sim c\xi/\Delta r$ from the point vortex model. This yields an acceleration $a \sim v/\Delta t$ and hence a prediction for the radiated energy, $E_c\sim c^3\xi^3/\Delta r^4$~\cite{kwon_sound_2021},
up to prefactors depending on the chemical potential $\mu$ and interaction strength $g$, and corrected by the finite core area of the vortices \cite{parker_controlled_2004, kwon_sound_2021, Vortex_rings_Roustekoski_2005}.

\begin{figure}
\centering
\includegraphics[scale = 0.5]{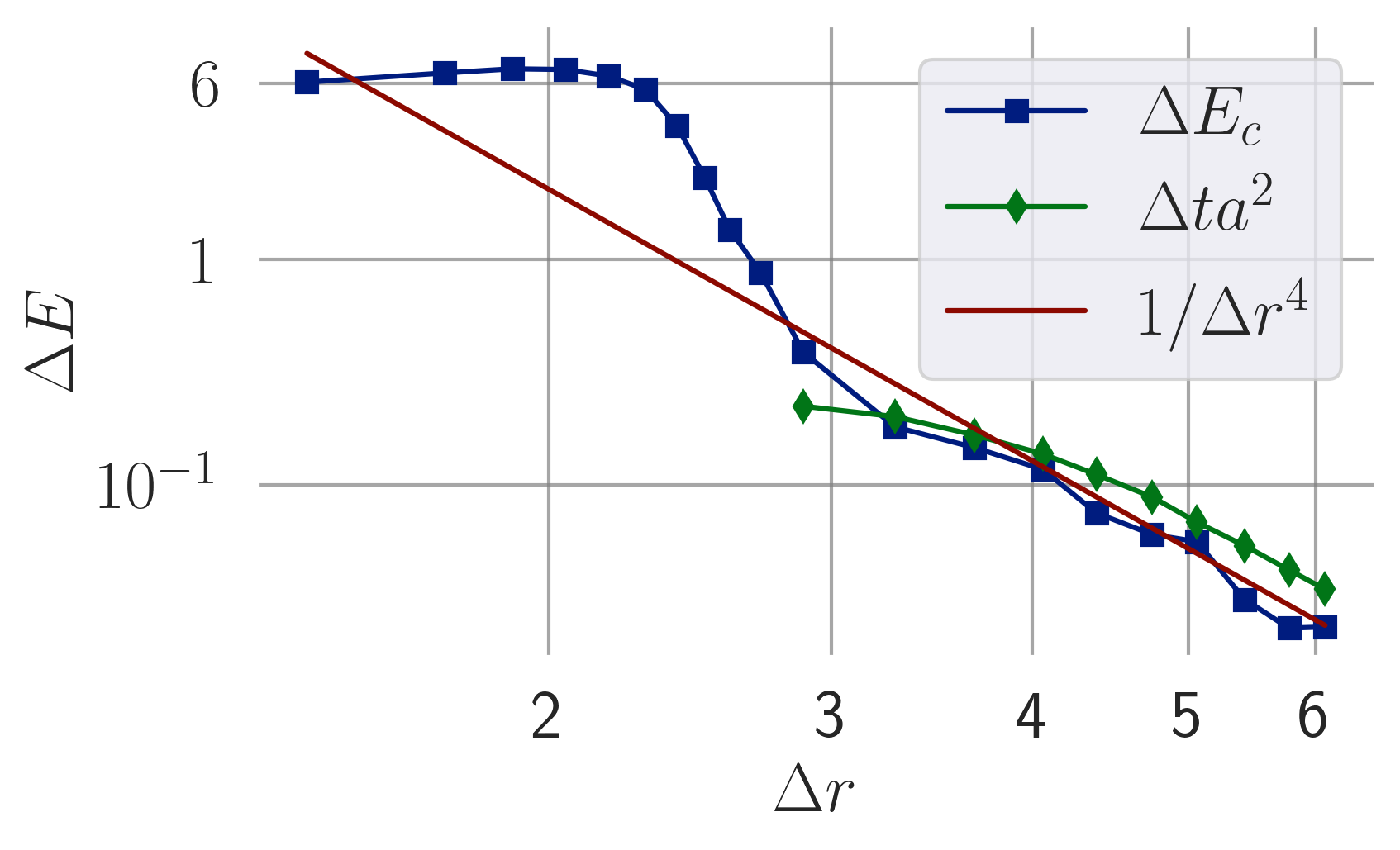}
\caption{Comparison of radiated compressible kinetic energy. Blue squares: numerical results. Green triangles: Larmor prediction. Red line: ideal head-on collision. Theoretical estimates include all four vortices and are scaled by a factor 30 for comparison.}
\label{fig:dipdip_energy2}
\end{figure}

The numerical results confirm this prediction (\autoref{fig:dipdip_energy2}). For $\Delta r \gg \xi$, the radiated compressible energy saturates at $\Delta E_c \sim 6\mu_0$, comparable in scale to the change in $E_i$. This value is much smaller than the typical per-vortex energy release observed in decaying turbulence, underscoring the distinct character of controlled dipole collisions compared to disordered vortex dynamics.

\section{Decaying quantum turbulence}\label{Sec:decay_turb}
Dipole collisions provide a useful reference for understanding energy transfer in decaying quantum turbulence. During a head-on dipole collision with $\Delta r<\Delta r_c$, the change in incompressible kinetic energy per vortex is small, i.e. $\Delta E_i/\Delta N_v\sim 1.5\mu_0$, much lower than typical per-vortex energy changes observed in turbulent simulations (\autoref{tab:qt_results}). This indicates that such collisions represent low-energy configurations of the vortices. In decaying turbulence, the most significant energy transfer occurs in processes that bring vortices into similar low-energy configurations, either through multiple collisions or through radiation induced by vortex acceleration. Consequently, vortex–antivortex annihilation events alone cannot account for the dominant transfer from incompressible to compressible modes.

\begin{table}[t]
\caption{Change in incompressible kinetic energy per vortex in decaying quantum turbulence with periodic boundary conditions (-) or confined to a disk (Trap). }
	\begin{center}
		\begin{tabular}{ccc}
        \cline{1-3}
		   $\gamma$  & $\Delta E_i/\Delta N_v$ [$\mu_0$]&  \\
         \cline{1-3}
			$0$  & 4.56 & - \\
   			$0$  & 4.03 & Trap \\
			$5\times10^{-4}$ & 5.29 &   -  \\
			$5\times10^{-4}$  & 5.26 &  Trap\\
			$10^{-2}$ & 6.41 &  - \\
			$10^{-2}$ & 6.43  & Trap \\    
		\end{tabular}
	\end{center}
    \label{tab:qt_results}
\end{table}

\begin{figure}
    \centering
    \includegraphics[width=\linewidth]{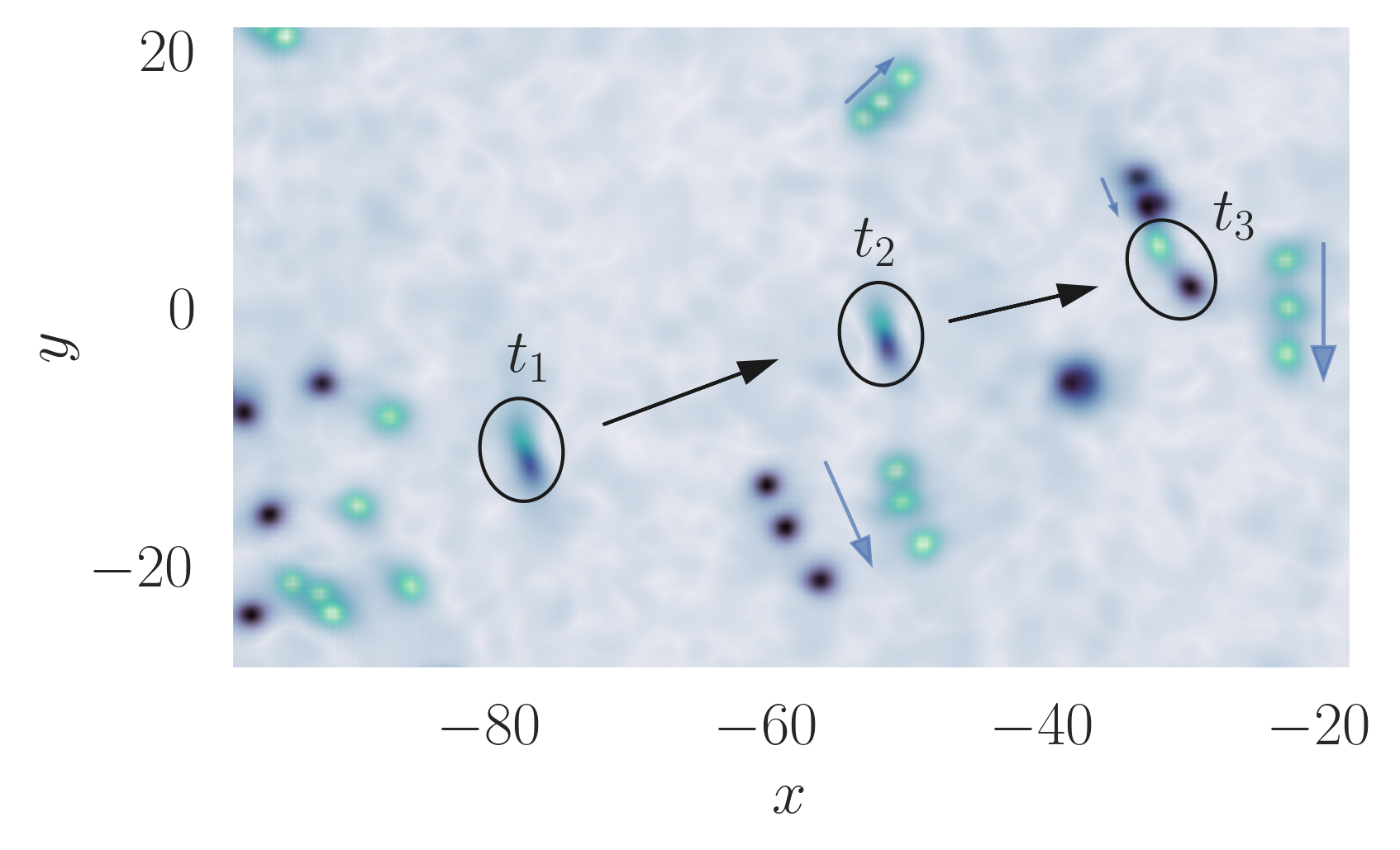}
    \caption{Three overlayed images of the $D$-field  at $t_1 = 550\tau$, $t_2 = 590\tau$ and $t_3 = 630\tau$. The regions of uniform condensate density are transparent, while waves and vortices remain visible. At $t_1$ and $t_2$, a solitary wave is moving from left to right. At $t_3$, the wave has split up into a dipole due to interaction with the surrounding vortices. The motion of the surrounding vortices is indicated by gray arrows.}
    \label{fig:turb_wave_splitting_D}
\end{figure}

Solitary waves play an important role in this energy redistribution. While in controlled collisions a solitary wave may persist after a single encounter, in turbulence it typically loses energy gradually over a series of interactions with multiple vortices. Rarely, a solitary wave can split into two or more regions of finite pseudo-vorticity, each carrying a separate phase singularity, as illustrated in \autoref{fig:turb_wave_splitting_D}. In this example, a solitary wave at $t= 550\tau$ and $t= 590\tau$ is intact, but by $t= 630\tau$ it has split into two dipole-like structures due to interactions with surrounding vortices. While such splitting events are uncommon and do not significantly affect the turbulence statistics, they reveal an additional layer of complexity in the phenomenology of the four-vortex mechanism.

\begin{figure}
    \centering
    \includegraphics[width=\linewidth]{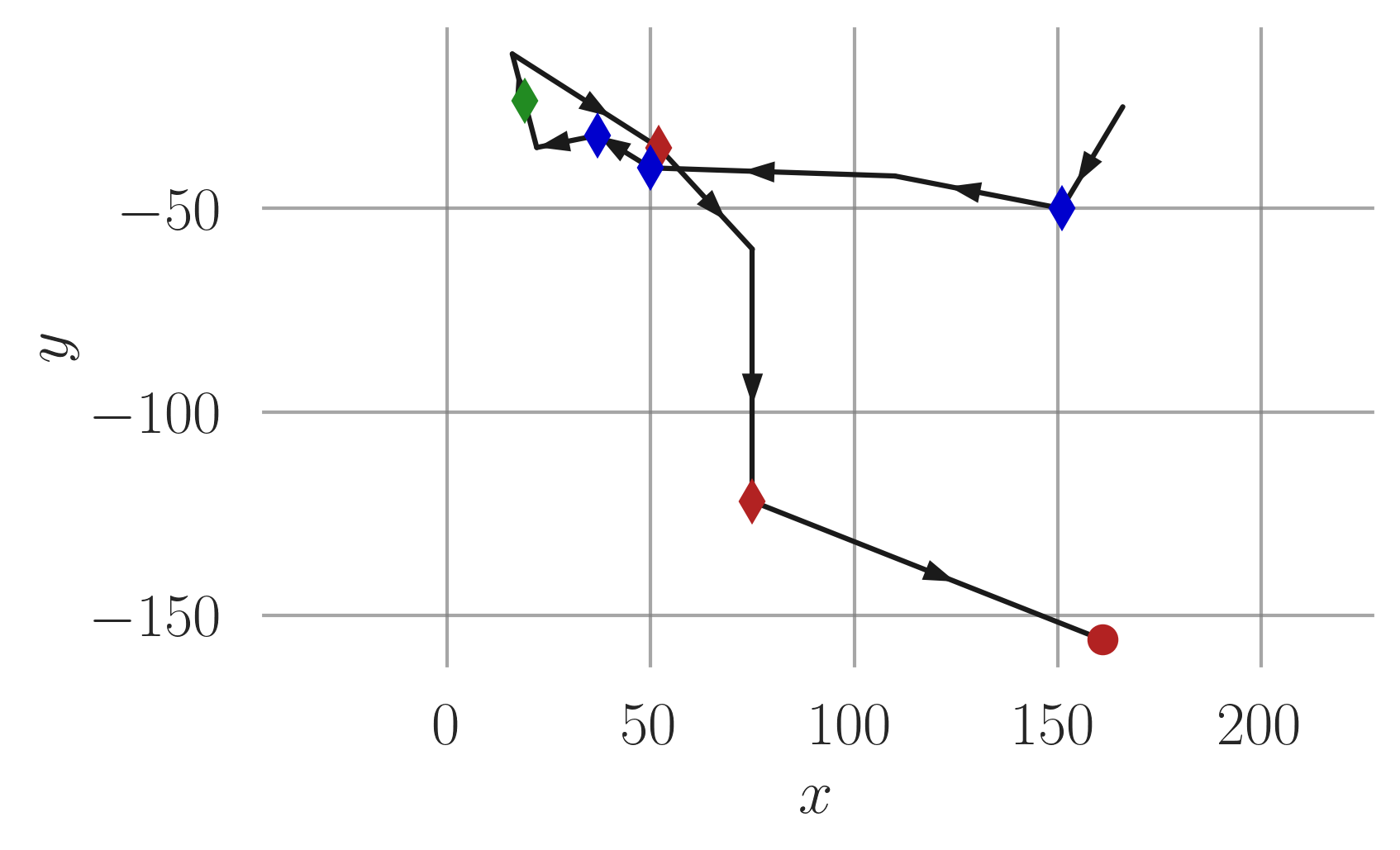}
    \caption{Illustration of the trajectory of a dipole and solitary wave. A small dipole starts in the top right corner and gradually shrinks through a series of exchange events, marked with the blue diamonds. At the exchange event by the green diamond the dipole collapses into a solitary wave. The red diamonds show the exchange events undergone by the solitary wave after collapse. The wave disperses into sound due to a collision with vortex at the red circle. The kinks in the trajectory that are not marked as exchanges are deflections due to interactions with vortices.}
    \label{fig:collision_series}
\end{figure}

\autoref{fig:collision_series} further illustrates the gradual dissipation of energy by a solitary wave in turbulence. A small dipole undergoes a series of exchange events (blue diamonds), collapses into a solitary wave (green diamond), and continues to interact with surrounding vortices through additional exchanges (red diamonds) before finally dispersing into sound (red circle). Kinks in the trajectory correspond to deflections caused by nearby vortices rather than direct exchanges.

\begin{figure}
    \centering
    \includegraphics[width=\linewidth]{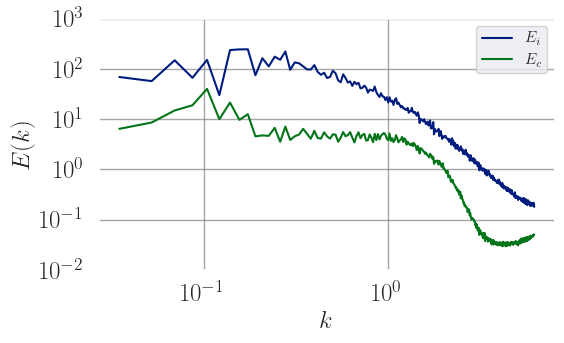}
    \caption{The spectra of the incompressible kinetic energy and the compressible kinetic energy at $t = 590 \tau$.}
    \label{fig:spectra}
\end{figure}
The kinetic energy spectra in \autoref{fig:spectra} provide complementary insights. The incompressible spectrum exhibits a clear energy cascade, reflecting the dynamics of vortices in the turbulent flow. The compressible spectrum, by contrast, shows equipartition at scales smaller than the vortex core, with a peak at larger scales associated with collective excitations such as density waves. Together, these observations demonstrate that in decaying turbulence, energy is redistributed through a combination of vortex interactions, radiation, and wave–vortex coupling, rather than through isolated annihilation events.

\section{Conclusions and discussion}\label{Sec:conclusion}

We have studied collision events between solitary waves and vortices, as well as head-on dipole collisions, and their contribution to the decaying quantum turbulence. 

Vortex-wave collisions, the interaction occurs as a topological exchanges, whereby the wave splits into a transient pair of phase singularities, one of which annihilates with the vortex, while the other replaces it. Whether the wave energy is irreversibly converted into sound depends on its initial strength. Low-energy waves disperse quickly, while high-energy waves resemble dipoles, persisting long enough to localize vorticity and nucleate vortices through interactions with other vortices or with the trap boundary. This indicates that repeated collisions are typically required for complete conversion of wave energy into sound.

For dipole–dipole collisions, we found that energy transfer from compressible to incompressible modes is mediated by the quantum energy. When non-annihilating vortices approach closely, the transfer is strongly influenced by density overlap, whereas at larger separations it follows the scaling of Larmor-type radiation, consistent with vortex acceleration. Comparing the incompressible energy released per dipole annihilation with that observed in decaying turbulence shows that dipole annihilation alone cannot account for the total transfer. Instead, the processes that reorganize vortices into low-energy, annihilation-prone configurations appear to play a more significant role than the annihilations themselves.

Together, our results indicate that the decay of quantum turbulence cannot be attributed solely to the energy released during individual annihilation events. Rather, it is driven by collective dynamics: repeated wave–vortex encounters, the restructuring of vortex configurations that facilitate annihilation, and the gradual conversion of energy into sound. A full understanding of turbulence decay therefore requires accounting not only for isolated interactions, but also for the mechanisms that organize and sustain them.

\begin{acknowledgments}
We thank Jonas Rønning, Pietro Massignan, Andrea Richaud and Daniel Pérez-Cruz for useful discussions and feedback. Vebjørn Øvereng acknowledges support by the Spanish Ministerio de Ciencia, Innovación y Universidades (grant PID2023-147469NB-C21, financed by MICIU/AEI/10.13039/501100011033 and FEDER-EU).
\end{acknowledgments}

\appendix

\section{Numerical details}{\label{Appendix}}
We use a disk shaped trapping potential given by 
\begin{equation*}
    V_{\text{Ext}} = \frac{V_0}{2}\left[1 - \tanh\left(\frac{r - R}{\chi}\right)\right],
\end{equation*}
where $R$ sets the radius of the disk, and $\chi$ the steepness. The system is initialized with the Thomas-Fermi wave function $|\Psi_{\text{TF}}(\mathbf{r})|^2 = (\mu - V_{\text{Ext}}(\mathbf{r}))/g$, before the vortices are imprinted by adding the vortex phase to the phase of the wave function 
$\Psi(\mathbf{r}) = \Psi_0(\mathbf{r})\exp{-i\sum_k\theta_k}$.
The phase of a vortex at $(x_k, y_k)$ is given by $\theta_k = q_k \arctan(\frac{x - x_k}{y-y_k})$, where $q_k=\pm 1$ is the vortex charge. 

To simulate the collision between a vortex and a solitary wave a vortex is imprinted at the center of an elliptical trap in a numerical domain of size $512\times256$, with a system size of $128\xi\times64\xi$, where $\xi = \hbar/\sqrt{m\mu}$ is the healing length of the condensate.
To produce a solitary wave, we imprint two vortices of opposite charge with separation in the range $\Delta r \in \left[0.8\xi, 1.8\xi\right]$. 
These will immediately collapse and form a solitary wave similar to the wave formed by a vortex annihilation, the abrupt change in density profile during the collapse also emits sound waves. To minimize the effects from this noise we add a non-zero damping along the boundary of the trap, this does not significantly influence the dynamics of the collision at the center of the trap, but it makes the position of the vortex easier to track. The wave is imprinted at a distance of $r_0 = 20\xi$ from the vortex.

The simulations of head-on collisions of vortex pairs are performed in a disk trap of radius $R = 29\xi$, with an initial distance between the vortex pairs of $20\xi$. The separation between the vortices in each pair is in the range $\Delta r \in \left[3\xi, 6 \xi\right]$.

To find the position of the phase slips within the vortices we first filter out regions where $|D|<\varepsilon_D$ and $\rho < \varepsilon_\rho$. Initially we set $\varepsilon_D = 0.1$ and $\varepsilon_\rho = 0.3$. Next we count the number of regions that satisfy these thresholds. Since there is an upper limit on the number of phase slips we expect to find, we can iteratively adjust $\varepsilon_\rho$ down until the number of regions located falls below the expected limit. To ensure that we include enough pixels around each phase slip we include all pixels within a radius of $\sqrt{2}\dd x$ and interpolate the superfluid density within this region. Finally, we find the minimum of the superfluid density within each region. This is our estimate for the position of the phase slips. 

The simulations of decaying quantum turbulence are performed on a numerical domain of $512\times512$. For three different values of the damping coefficient, $\gamma \in \{0, 5\times10^{-4}, 10^{-2}\}$, we perform one simulation with periodic boundary conditions, and one where the condensate is trapped in a disk potential. For the simulations with periodic boundary conditions we start out with $\sim 100$ vortices in a random configuration, for the disk potential the initial vortex number is $\sim 45$. We average over 8 random initial vortex configurations for each of the six simulations.

\bibliographystyle{unsrt} 
\bibliography{references} 

\end{document}